\documentclass[12pt,english]{extarticle}
\usepackage[T1]{fontenc}
\usepackage[utf8]{inputenc}
\usepackage[a4paper]{geometry}
\usepackage{amsmath}
\usepackage{amssymb}
\usepackage{setspace}
\usepackage{esint}

\makeatletter
\newenvironment{lyxlist}[1]
{\begin{list}{}
{\settowidth{\labelwidth}{#1}
 \setlength{\leftmargin}{\labelwidth}
 \addtolength{\leftmargin}{\labelsep}
 }}
{\end{list}}

\usepackage{ifxetex}\ifxetex
\usepackage{fontspec}
\else
\fi

\usepackage{fixltx2e}\usepackage{fancyhdr}

\usepackage{babel}

\makeatother

\usepackage{babel}
\begin{document}

\title{\textbf{Bohmian Zitterbewegung }}

\author{Giuseppe Raguní\thanks{UCAM University of Murcia, Spain - graguni@ucam.edu}}

\date{27-may-2021}
\maketitle
\begin{abstract}
A new Bohmian quantum-relativistic model, in which from the Klein-Gordon
equation a generalization of the standard \emph{Zitterbewegung} arises,
is explored. It is obtained by introducing a new independent time
parameter, whose relative motions are not directly observable but
cause the quantum uncertainties of the observables. Unlike Bohm's
original theory, the quantum potential does not affect the observable
motion, as for a normal external potential, but it only determines
that one relative to the new time variable, of which the \emph{Zitterbewegung}
of a free particle is an example. The model also involves a relativistic
revision of the uncertainty principle. \end{abstract}
\begin{lyxlist}{00.00.0000}
\item [{KEYWORDS:}] Quantum Mechanics; de Broglie-Bohm interpretation;
Relativity; Klein-Gordon equation; \emph{Zitterbewegung}; Uncertainty
principle verification; Time 
\end{lyxlist}

\section{Introduction}

In the de Broglie-Bohm interpretation of the non-relativistic quantum
mechanics, determinism is recovered with the introduction of a specific
quantum potential {[}1-5{]}. The difficulties or even the very existence
of a coherent relativistic generalization of this approach, have been
questioned by many researchers {[}7-15{]}. However, one of the main
problems, namely the compatibility of the \emph{non-locality} with
the Lorentz covariance, has been judged solvable by some authors {[}16-23{]}.

In this work we begin by showing that a relativistic generalization
of the original de Broglie-Bohm theory is unsatisfactory as regards
the very continuity of the quantum potential between the relativistic
and non-relativistic cases. Another serious limitation of the current
Bohmian relativistic theories is the inability to derive the \emph{Zitterbewegung}
{[}25-26{]} of a free particle. After, we show that both problems
can be solved by introducing a new independent time parameter, whose
relative motions - such as the \emph{Zitterbewegung} - are not observable
but are the cause of the uncertainties of the quantum observables.
The quantum potential, unlike a normal external potential, does not
influence the observable motion but only affects that one relative
to the new time parameter. It turns out that the \emph{Zitterbewegung}
of a free particle is an example of this motion because it originates
from the quantum potential. A relativist\emph{ }revision of the uncertainty
principle also derives from the studied model.

The paper is organized as follows: in section 2 we recall the solutions
of the de Broglie-Bohm model for a non-relativistic free particle.
In the following 3 and 4 we show the problems of every possible relativistic
generalization of these motions, adopting the original Bohm's model.
In section 5 the new Bohmian model is defined and in section 6 it
is applied to study the free particle, finding a generalization of
both the standard \emph{Zitterbewegung} and the uncertainty principle.
The last section 7 collects final considerations.

\section{Bohmian solutions for a non-relativistic free particle}

The de Broglie-Bohm approach {[}1-5{]} consists in replacing the wave
function:

\begin{equation}
\psi(\vec{r},t)=R(\vec{r},t)\,e^{\frac{i}{\hbar}S(\vec{r},t)}
\end{equation}
where \emph{R$>0$} and \emph{S }are real functions, into the Schrödinger
equation:

\begin{equation}
i\hbar\frac{\partial\psi}{\partial t}=-\frac{\hbar{}^{2}}{2m}\,\nabla{}^{2}\psi+V\,\psi
\end{equation}
Two equations are obtained:

\begin{equation}
\frac{\partial S}{\partial t}+\frac{(\vec{\nabla}S){}^{2}}{2m}\,+V-\frac{\hbar{}^{2}}{2m}\,\frac{\nabla{}^{2}R}{R}=0
\end{equation}

\begin{equation}
\frac{\partial R^{2}}{\partial t}+\vec{\nabla}\cdot(R^{2}\frac{\vec{\nabla}S}{m})=0
\end{equation}
Now, by identifying \emph{S} with the Hamilton's principal function,
we have $\vec{\nabla}S=\vec{p}$ , and eq. (3) represents the Hamilton-Jacobi
equation with Hamiltonian:

\begin{equation}
H=\frac{p{}^{2}}{2m}+V-\frac{\hbar{}^{2}}{2m}\,\frac{\nabla{}^{2}R}{R}
\end{equation}
where the last addend represents a potential energy that takes into
account the quantum effects:

\begin{equation}
V_{Q}\equiv-\frac{\hbar{}^{2}}{2m}\,\frac{\nabla{}^{2}R}{R}
\end{equation}
Eq. (4), on the other hand, shows that \emph{R}$^{2}$, and \emph{R}
itself, is a wave that follows the particle with velocity $\vec{v}$.

According to the Bohm's original quantum interpretation, the motion
equation of the particle is:

\begin{equation}
\frac{d\vec{p}}{dt}=-\vec{\nabla}V-\vec{\nabla}V_{Q}
\end{equation}
In the case of a free particle (\emph{V }= 0), Bohm showed the possibility
of non-stationary solutions, with $\vec{p}$, $V_{Q}$ and \emph{H}
variable in time {[}6{]}. However, there are also stationary solutions.
To find them, let's assume that $\vec{p}$ is constant, making coincide
with $\vec{x}$ the rectilinear trajectory. Let's introduce the variable
$\ell\equiv x-v\,t$, denoting with a dot the derivative with respect
to $\ell$. Since $\hat{x}=\hat{v}$ , we obtain $\vec{\nabla}R=\dot{R\,}(\ell)\,\hat{v}$
and, applying the divergence operator: 
\begin{equation}
\nabla^{2}R=\ddot{R}(\ell)
\end{equation}
Substituting into the eq. (6), we obtain:

\begin{equation}
-\frac{2m}{\hbar^{2}}\,V_{Q}=\frac{\ddot{R}}{R}
\end{equation}
The accomplishment of the eq. (7) requires that $V_{Q}$ be constant.
Placing $k\equiv-\frac{2m}{\hbar^{2}}\,V_{Q}$, for \emph{k}>0, i.e.
$V_{Q}$ < 0, eq. (9) is satisfied for:

\begin{equation}
R(\ell)=A\,e^{\sqrt{k}\,\ell}+B\,e^{-\sqrt{k\,}\ell}
\end{equation}
with \emph{A} and \emph{B} arbitrary constants. Nonetheless, these
solutions are physically unacceptable because they diverge for $\mid\ell\mid\rightarrow\infty$. 

For \emph{k}<0, i.e. $V_{Q}$ > 0 , we get instead:

\begin{equation}
R(\vec{r},t)=\mid A\,cos\frac{\sqrt{2mV_{Q}}}{\hbar}(x-v\,t)\mid
\end{equation}
where \emph{A} is an arbitrary constant, and we have imposed, just
for simplicity, the simmetry $R\,(\ell)=R\,(-\ell)$, assuming that
at the initial instant, \emph{t}=0, the particle is in the origin.
The solutions (11), although not normalizable, are limited and hence
can fit in a satisfactory wave packet description. 

We have therefore obtained solutions where $\vec{p}$, $V_{Q}$ and
hence \emph{H} are constant, then close to that one of the standard
quantum mechanics. But here \emph{R} depends on space and time.

In the following sections, we will seek a relativistic generalization
of the previous result, using the Klein-Gordon equation and limiting
ourselves to consider only non-negative values for the Hamiltonian.

\section{First attempt of relativistic generalization}

In reference to eq. (1), let's interpret \emph{S} as the Hamilton's
principal function, by writing:

\begin{equation}
\vec{\nabla}S=\vec{p}=m\,\gamma\,\vec{v}
\end{equation}

\begin{equation}
\frac{\partial S}{\partial t}=-H=-m\,\gamma\,c^{2}-V_{Q}
\end{equation}
where \emph{m} is the rest mass, $\gamma$ the Lorentz factor and
we have introduced, for generality, a quantum potential $V_{Q}$ to
be determinated.

We will start with the previous stationary conditions: $\vec{p}$,
$V_{Q}$ and \emph{H} constant.

From eq. (1), taking the gradient, then the divergence, and finally
dividing by $\psi$, we obtain:

\begin{equation}
\frac{\nabla{}^{2}\psi}{\psi}=\frac{\nabla{}^{2}R}{R}+2\,\frac{i}{\hbar}\,\frac{\vec{\nabla}R}{R}\,\cdot\,\vec{p}-\frac{p^{2}}{\hbar{}^{2}}
\end{equation}
being $\vec{\nabla}$·$\vec{v}$ = 0. The comparison with the Klein-Gordon
equation:

\begin{equation}
\nabla{}^{2}\psi=\frac{m{}^{2}c^{2}}{\hbar{}^{2}}\,\psi+\frac{1}{c^{2}}\,\frac{\partial^{2}\psi}{\partial t^{2}}
\end{equation}
provides:

\begin{equation}
\frac{\nabla{}^{2}R}{R}+2\,\frac{i}{\hbar}\,\frac{\vec{\nabla}R}{R}\,\cdot\,\vec{p}=\frac{m{}^{2}\gamma^{2}c^{2}}{\hbar{}^{2}}+\frac{1}{c^{2}\psi}\,\frac{\partial^{2}\psi}{\partial t^{2}}
\end{equation}
where we applied the identity $p^{2}+m^{2}c^{2}=m^{2}\gamma^{2}c^{2}$.

On the other hand, by deriving eq. (1) successively with respect to
time, we obtain:

\begin{equation}
\frac{\partial\psi}{\partial t}=\frac{\partial R}{\partial t}\,e^{\frac{i}{\hbar}S}+\frac{i}{\hbar}\,\psi\,(-H)
\end{equation}

\begin{equation}
\frac{\partial^{2}\psi}{\partial t^{2}}=\frac{\partial^{2}R}{\partial t^{2}}\,e^{\frac{i}{\hbar}S}+\frac{i}{\hbar}\,\frac{\partial R}{\partial t}\,e^{\frac{i}{\hbar}S}(-H)+\frac{i}{\hbar}\,(-H)\,\frac{\partial\psi}{\partial t}
\end{equation}
from which it can be inferred:

\begin{equation}
\frac{1}{\psi}\frac{\partial^{2}\psi}{\partial t^{2}}=\frac{1}{R}\frac{\partial^{2}R}{\partial t^{2}}-2\,\frac{i}{\hbar}\,H\frac{1}{R}\frac{\partial R}{\partial t}-\frac{H^{2}}{\hbar{}^{2}}
\end{equation}
By substituting eq. (19) into eq. (16) and matching the real and imaginary
parts, we obtain:

\begin{equation}
H^{2}=m^{2}\gamma^{2}c^{4}+\hbar{}^{2}(\frac{1}{R}\frac{\partial^{2}R}{\partial t^{2}}-c^{2}\frac{\nabla{}^{2}R}{R})
\end{equation}

\begin{equation}
\vec{p}\,\cdot\vec{\nabla}R+\frac{H}{c^{2}}\,\frac{\partial R}{\partial t}=0
\end{equation}
Now we suppose that \emph{R} is a generic wave that follows the particle
with velocity $\vec{v}$. By limiting the study to the rectilinear
trajectory, which we make coincide with $\vec{x}$ , let's introduce
the variable $\ell\equiv x-v\,t$, obtaining as before:

\begin{spacing}{0.4}
 
\begin{equation}
\vec{\nabla}R=\dot{R\,}(\ell)\,\hat{v}
\end{equation}

\end{spacing}

\begin{singlespace}
\begin{equation}
\nabla^{2}R=\ddot{R}(\ell)
\end{equation}
The derivatives with respect to the time of \emph{R} give:
\end{singlespace}

\begin{equation}
\frac{\partial R}{\partial t}=-\dot{R\,}(\ell)\,v
\end{equation}

\begin{equation}
\frac{\partial^{2}R}{\partial t^{2}}=v^{2}\ddot{R}(\ell)
\end{equation}
By substituting the last four equations in eqs. (20) and (21), we
finally obtain:

\begin{equation}
H^{2}=m^{2}\gamma^{2}c^{4}+\hbar{}^{2}(v^{2}-c^{2})\frac{\ddot{R}(\ell)}{R}
\end{equation}

\begin{equation}
v\,\dot{R\,}(\ell)\,(m\,\gamma\,c^{2}-H)=0
\end{equation}

Since we are looking for non-trivial solutions, we will assume that
\emph{m$\gamma$} and \emph{v} are non-zero; moreover that \emph{R}
is not constant, as we found in the non-relativistic case. But in
such hypotheses, eq. (27) implies $H=m\,\gamma\,c^{2}$ and therefore
$V_{Q}=0$, while in the non-relativistic case we have seen that it
is admitted that it is a non-zero constant. Now, it is true that this
does not imply any absurdity, since\emph{ H} is defined up to an arbitrary
additive constant; however, the fact remains that in the context of
Bohm's interpretation one cannot assign to the quantum potential a
straightforward continuity between the relativistic and non-relativistic
cases, and this is not very satisfactory.

For this reason it seems appropriate to look for an alternative.

\section{No alternative in the Bohm's standard interpretation}

Keeping the hypothesis that \emph{H} is a constant of motion, let's
try to introduce a time dependence for the momentum of the particle.
Eqs. (20) and (21) do not change. For \emph{R}, we will continue to
assume that it is a wave following the particle, and therefore an
arbitrary function of $\ell\equiv x-\intop_{0}^{t}v(\xi)\,d\xi$ ;
but eqs. (22), (24) and, consequently, (27) remain unchanged, so one
still gets the unsatisfactory solution $V_{Q}=0$.

We therefore reach a first conclusion: if there is a more satisfactory
relativistic generalization of the Bohmian quantum mechanics, in it
\emph{the Hamiltonian cannot be a constant of motion} \footnote{This is possible if $V_{Q}$ is not a \emph{generalized potential},
see e.g. {[}24{]}.}\emph{. }

If \emph{H} is a function of time, the case in which the moment of
the particle remains constant is still impossible if we keep the hypothesis
that\emph{ R} is an arbitrary function of $\ell\equiv x-v\,t$. In
fact, based on eq. (26), $V_{Q}$ is a function of $\frac{\ddot{R}(\ell)}{R}$
and therefore also a function of $\ell$. Then, from the equation
of motion:

\begin{equation}
\frac{d\vec{p}}{dt}=-\vec{\nabla}V_{Q}=-\dot{V}_{Q}(\ell)\,\hat{v}=0
\end{equation}
it follows that $V_{Q}$ is constant. But then, from eq. (13), \emph{H}
should also be constant.

Finally, let us explore the possibility that \emph{H} and $p$ are
both variable, with $\hat{v}$ constant. By assuming for \emph{R}
an arbitrary function of $\ell\equiv x-\intop_{0}^{t}v(\xi)\,d\xi$,
eqs. (22), (23) and (24) do not change, while eq. (25) transforms
into:

\begin{equation}
\frac{\partial^{2}R}{\partial t^{2}}=v^{2}\ddot{R}(\ell)-\dot{R\,}(\ell)\,v'(t)
\end{equation}
where the apostrophe indicates a derivative with respect to time.
Substituting in eq. (20) we obtain:

\begin{equation}
H=\sqrt{m^{2}\gamma^{2}c^{4}-\hbar{}^{2}\frac{c^{2}}{\gamma^{2}}\frac{\ddot{R}}{R}-\hbar{}^{2}v'\frac{\dot{R}}{R}}
\end{equation}

On the other hand, the time dependence for \emph{H} implies the new
addend $-\frac{i}{\hbar}\frac{\partial H}{\partial t}$ to the second
member of eq. (19), so, in place of (27), we now have:

\begin{equation}
v\,\frac{\dot{R}}{R}\,(m\,\gamma\,c^{2}-H)+\frac{1}{2}\frac{\partial H}{\partial t}=0
\end{equation}
Substituting eq. (30) into eq. (31), being $\gamma'=\frac{\gamma^{3}}{c^{2}}\,v$$\,v'$,
we arrive at the following expression:

\begin{equation}
(H-m\gamma c^{2}-\frac{\hbar{}^{2}v''}{4m\gamma c^{2}v})\frac{4\dot{R}}{mv'R}+\frac{3\hbar{}^{2}}{m^{2}\gamma c^{2}}(\frac{\dot{R}^{2}}{R^{2}}+\frac{\ddot{R}}{R})+\frac{\hbar{}^{2}}{m^{2}\gamma^{3}v'}(\frac{3\ddot{R}\dot{R}}{R^{2}}+\frac{\dddot{R}}{R})+2\gamma^{3}=0
\end{equation}
where we admit $v'\neq0$, in addition to the previous hypotheses
of non-triviality. Eq. (32) should hold whatever the rest mass \emph{m}
is, but it is clear that this is impossible: when \emph{m} increases,
all the addends become small, except the last one which remains a
non-zero number.

This latter failure exhausts the possibilities, for Bohm's original
quantum model, of the existence in {[}1+1{]} dimension of a relativistic
generalization more satisfactory than that one we found in the previous
section.

\section{The new Bohmian relativistic model}

The attempt with variable \emph{H} and constant moment failed due
to the Bohmian equation (28); one could then think of referring it
to a new, independent, intrinsic moment $\vec{p}_{i}$, variable over
time:

\begin{equation}
\frac{d\vec{p}_{i}}{dt}=-\vec{\nabla}V_{Q}
\end{equation}
This motion could coincide with the \emph{Zitterbewegung} of a free
particle, the famous sinusoidal motion of amplitude $\frac{\hbar}{2m\gamma c}\,$
and angular frequency $\frac{2m\gamma c{}^{2}}{\hbar}\,$ in any direction
{[}25-26{]}. The \emph{Zitterbewegung} emerges in the Heisenberg's
\emph{picture} and the debate on its exact interpretation - and even
observability - is not yet resolved {[}27-40{]}. To find this peculiar
movement in a deterministic approach could clarify its nature.

First, the idea given by (33) does not work. In the sketched picture,
the momentum of the particle would be:

\begin{equation}
\vec{p}\equiv\vec{p}_{o}+\vec{p}_{i}
\end{equation}
where $\vec{p}_{o}$ is the constant moment (actually observed in
mean value), to be substituted for $\vec{p}$ in all the equations
written above, except those (7) and (28). In the one-dimensional simplification,
a wave following the particle would then be an arbitrary function
of $\ell\equiv x-v_{o}\,t-\intop_{0}^{t}v_{i}(\xi)\,d\xi$; however,
we have seen that for \emph{R} we must consider a dependence only
on $x-v_{o}t$ to avoid getting the impossible eq. (32) again. But
it is obvious that if \emph{R} does not also depend on $v_{i}$, eq.
(33) is unable to account for the intrinsic motion. The idea of neglecting
the intrinsic movement in derivatives with respect to time, due to
the fact that it consists of very rapid oscillations, could work in
many cases as an approximation but strictly, as we have seen at the
end of the previous section, it is incorrect in {[}1+1{]} dimension.

The solution here explored is the introduction of a \emph{new independent
time parameter} $\tau$ for the spatial coordinates (\emph{X},\emph{Y},\emph{Z})of
the particle. That is, we admit that they depend not only on commonly
experienced time \emph{t}, but also on another, totally independent,
time parameter $\tau$. By referring, for simplicity, only to \emph{$X(t,\tau)$},
let's introduce two velocities:

\begin{equation}
v_{o}\equiv\frac{\partial X}{\partial t}\,
\end{equation}

\begin{equation}
v_{ix}\equiv\frac{\partial X}{\partial\tau}\,
\end{equation}
having so:

\begin{equation}
dX=v_{o}dt+v_{ix}d\tau
\end{equation}
If we suppose that $v_{o}$ and $v_{i}$ depend, respectively, only
on $t$ and $\tau$ , by integrating, we get so:

\begin{equation}
X(t,\tau)=X(0,0)+\intop_{0}^{t}v_{o}(\xi)\,d\xi+\intop_{0}^{\tau}v_{ix}(\zeta)\,d\zeta
\end{equation}

Due to the independence of $\tau$, we have to recognize that the
intrinsic motion is not directly observable and is perceived as \emph{instantaneous}
with respect to time \emph{t}. This does not at all mean, of course,
that it does not have dramatic effects on the measurements of the
observable quantities. As for the position, the eq. (38) allows to
the particle, observed in its trajectory as a function of \emph{t},
to jump instantly from one point of space to another, in principle
arbitrarily distant, through the time dimension $\tau$. Actually,
quantum mechanics is not new to this type of non-local peculiarity.
In the Feynman's \textquotedbl{}sum of histories\textquotedbl{} interpretation,
for example, the physical characteristics of a displacement from \emph{A}
to \emph{B}, for a particle or photon, can be explained correctly
only by admitting that it has traveled simultaneously all the possible
trajectories from \emph{A} to \emph{B} \footnote{Doing everything that was possible to do, such as emitting or absorbing
an arbitrary number of photons (if it is an electron) and interacting
with every other particle in all possible ways, even exceeding \emph{c}.}. In the rear, to make the described uncertainty in accordance with
quantum mechanics, just recognize that the $\vec{r}(\tau)$ function
should vary where $R^{2}$ is not null and we have, therefore, some
probability to find the particle. In our case, in which the free particle
is described by a single wave, we will have that $\vec{r}(\tau)$
has a Compton length order maximum amplitude, since this value represents
the minimum spatial location. If, however, the particle is described
by a wave packet, then the maximum excursion for the position will
be of the order of $\lambda=\frac{\hbar}{p}$. From this point of
view, hence, the motion in $\tau$ is\emph{ direct cause} of the quantum
uncertainties.

In general, a generic wave that follows the particle will be a function
of $\vec{\ell}\equiv\vec{r}-\intop_{0}^{t}\vec{v}_{o}(\xi)\,d\xi\,-\intop_{0}^{\tau}\vec{v}_{ix}(\zeta)\,d\zeta$,
setting: $\vec{r}(0,0)=0$. The variable $\vec{\ell}$ includes the
dependence on $t$ and $\tau$, having $\frac{\partial\vec{\ell}}{\partial t}=-\vec{v}_{o}$
and:

\begin{equation}
\frac{\partial\vec{\ell}}{\partial\tau}=-\vec{v}_{ix}
\end{equation}

For the intrinsic motion we will assume that the \emph{Restricted
Relativity does not hold}. This condition is consistent with the fact
that, in this peculiar motion, the mass of the particle does not undergo
any relativistic increase - other than that one due to the observable
velocity - despite it reaches the speed \emph{c}. That aside, the
motion in $\tau$ obeys the classical laws of motion, such as conservation
of energy and momentum. Considering that the mass involved in the
generic intrinsic motion is $m\gamma_{o}$, energy conservation in
$\tau$ reads:

\begin{equation}
V_{Q}+\frac{1}{2}m\gamma_{o}v_{i}^{2}\equiv E_{Q}
\end{equation}
independent on $\tau$. It represents the \emph{purely quantum} total
energy.

On the other hand, in presence of a conservative potential $V$ ,
the following energy will be constant:

\begin{equation}
m\gamma_{o}c^{2}+V=const\equiv E_{C}
\end{equation}
that is the \emph{classic} total energy.

Taking the $\frac{\partial}{\partial\tau}$ of eq. (40), being $\frac{\partial V_{Q}}{\partial\tau}=-\vec{\nabla}V_{Q}\cdot\vec{v}_{i}$
as wave in $\tau$ , one finds: 

\begin{equation}
-m\gamma_{o}\frac{d\vec{v}_{i}}{d\tau}=m\gamma_{o}\frac{\partial^{2}\vec{\ell}}{\partial\tau^{2}}=-\vec{\nabla}V_{Q}(\vec{\ell})
\end{equation}

On the other hand, from eq. (41), one gets:

\begin{equation}
\frac{d\vec{p}_{o}}{dt}=-\vec{\nabla}V
\end{equation}
So, (42) and (43) are the equations of motion to replace Bohm's original
one (7)\footnote{This latter is not valid because it derives from admitting $\frac{dV_{Q}}{dt}=\vec{\nabla}V_{Q}\cdot\vec{v}_{o}$,
but, considering $V_{Q}$\emph{ }as a wave, now we have  $\frac{dV_{Q}}{dt}=\vec{\nabla}V_{Q}\cdot\vec{v}_{o}+\frac{\partial V_{Q}}{\partial t}=0$.}.

\medskip{}

The new Bohmian model can thus be so summarized: 
\begin{enumerate}
\item The spatial coordinates of the particle have to vary as a function
of two independent temporal parameters, \emph{t }and $\tau$. Motion
in \emph{$\tau$ }is ominidirectional, not directly observable and
responsible for quantum uncertainties.
\item The particle is represented by the wavefunction $\psi(\vec{r},t,\tau)=R(\vec{r},t,\tau)\,e^{\frac{i}{\hbar}S(\vec{r},t,\tau)}$,
where $\vec{\nabla}S=\vec{p}_{o}=m\gamma_{o}\vec{v}_{o}$ and $\frac{\partial S}{\partial t}=-H=-m\gamma_{o}c^{2}-V_{Q}-V$.
It obeys the Klein-Gordon equation, generalized for presence of potentials.
\item The wave function module\emph{ R} - and therefore also $V_{Q}$ and
\emph{$H$} - must be considered as wave following the particle, i.
e. as function of $\vec{\ell}$. Consequentially, the Bohmian law
of motion (7) must be replaced by eq. (43) for motion in \emph{t}
and by eq. (42) for motion in $\tau$, for which the Restricted Relativity
is not valid.
\end{enumerate}
We also observe that taking the gradient of $\psi$ and applying $\vec{\nabla}S=\vec{p}_{c}$
one can easily find the guidance equation:
\[
\vec{p}_{c}=\hbar\,Im\,(\frac{\vec{\nabla}\psi}{\psi})
\]

\section{\emph{Zitterbewegung }in the new Bohmian model}

Let's study finally a free particle in the newly introduced Bohmian
model. From eq. (43) we deduce that $\vec{p}_{o}$ is constant. The
equations referring to\emph{ t} time, when \emph{H} is variable, were
previously found:

\begin{equation}
H=\sqrt{m^{2}\gamma_{o}^{2}c^{4}-\hbar{}^{2}\frac{c^{2}}{\gamma_{o}^{2}}\frac{\ddot{R}}{R}}
\end{equation}

\begin{equation}
v_{o}\,\frac{\dot{R}}{R}\,(m\,\gamma_{o}\,c^{2}-H)+\frac{1}{2}\frac{\partial H}{\partial t}=0
\end{equation}
where we remember that $v_{o}$ is the observed (on average) velocity.

Placing $\lambda_{r}^{2}\equiv\frac{\hbar{}^{2}}{m^{2}c^{2}\gamma_{o}^{4}}$
- the square of the relativistic Compton length divided again by a
Lorentz factor - eqs. (44) and (45) become:

\begin{equation}
H=m\gamma_{o}c^{2}\,\sqrt{1-\lambda_{r}^{2}\,\frac{\ddot{R}}{R}}
\end{equation}

\begin{equation}
v_{o}\,\frac{\dot{R}}{R}\,(\frac{H}{m\gamma_{o}\,c^{2}}-\frac{H^{2}}{m^{2}\gamma_{o}^{2}c^{4}})+\frac{1}{4\,m^{2}\gamma_{o}^{2}c^{4}}\frac{\partial H^{2}}{\partial t}=0
\end{equation}

By squaring eq. (46) and substituting in eq. (47) we get:

\begin{equation}
v_{o}\,\frac{\dot{R}}{R}\,(\sqrt{\beta}-\beta)=-\frac{1}{4}\frac{\partial\beta}{\partial t}
\end{equation}
where $\beta(\ell)\equiv1-\lambda_{r}^{2}\,\frac{\ddot{R}}{R}$. Since
it is $\frac{\partial\beta(\ell)}{\partial t}=-\dot{\beta\,}(\ell)\,v_{o}$,
we obtain:

\begin{equation}
\frac{\dot{R\,}(\ell)}{R}\,=\frac{1}{4}\,\frac{\dot{\beta}}{\sqrt{\beta}-\beta}
\end{equation}

Equation (49) can be integrated member by member after multiplying
by $d\ell$. After easy steps, one finds:

\begin{equation}
1+\frac{c_{1}}{R^{2}}=\sqrt{\beta}
\end{equation}
where $c_{1}$ is an arbitrary constant. Recall that we are limiting
ourselves to consider only non-negative values for the Hamiltonian,
so discarding the possibility of having $-\sqrt{\beta}$ instead of
$\sqrt{\beta}$ in eqs. (48-50). On the basis of eq. (46) we therefore
obtain:

\begin{equation}
H=m\gamma_{o}c^{2}\,(1+\frac{c_{1}}{R^{2}})
\end{equation}
and so the quantum potential is: $V_{Q}=m\gamma_{o}c^{2}\,\frac{c_{1}}{R^{2}}$.
Compatibility with the stationary non-relativistic case requires $c_{1}$
be positive, as we have seen. By squaring eq. (50) and substituting
$\beta$, we find this differential equation for \emph{R}:

\begin{equation}
(1+\frac{c_{1}}{R^{2}})^{2}=1-\lambda_{r}^{2}\,\frac{\ddot{R}}{R}
\end{equation}
Recalling that \emph{R}>0, a first integration provides:

\begin{equation}
\dot{R}=\pm\frac{1}{\lambda_{r}}\,\sqrt{\frac{c_{1}^{2}}{R^{2}}-4c_{1}ln\,R+c_{2}}
\end{equation}
where $c_{2}$ is an arbitrary constant. Equation (53) is compatible
with a $R(\ell)$ even and with a maximum in the origin, which we
denote by $R_{M}$: so, we will have sign - for $\ell>0$ and + for
$\ell<0$. In this way we determine $c_{2}$, getting:

\begin{equation}
\dot{R}=\pm\frac{\sqrt{c_{1}}}{\lambda_{r}}\,\sqrt{\frac{c_{1}}{R^{2}}-\frac{c_{1}}{R_{M}^{2}}+4\,ln\,\frac{R_{M}}{R}}
\end{equation}
Placing $f\equiv\frac{c_{1}}{R_{M}^{2}}>0$, the quantum potential
is rewritten:

\begin{equation}
V_{Q}=m\gamma_{o}c^{2}f\,\frac{R_{M}^{2}}{R^{2}}
\end{equation}
and its minimum value is in the origin, where $R=R_{M}$ ; it is:
$V_{Qm}=m\gamma_{o}c^{2}f$. \emph{R} is determined by integrating
eq. (54):

\begin{equation}
\int\frac{dR}{\sqrt{f(\frac{R_{M}^{2}}{R^{2}}-1)+4\,ln\,\frac{R_{M}}{R}}}=-\frac{R_{M}\sqrt{f}}{\lambda_{r}}\mid\ell\mid+c_{3}
\end{equation}
which cannot be simplified by means of standard functions.

In a small neighborhood of the origin, i.e. for $R\rightarrow R_{M}$,
eq. (56) gives for \emph{R} a quadratic dependence on $\ell$. Here,
in fact, the rooting at first member is approximated by $2(f+2)(1-\frac{R}{R_{M}})$,
so by integrating we have:

\begin{equation}
\sqrt{1-\frac{R}{R_{M}}}\simeq\frac{\sqrt{f(f+2)}}{\sqrt{2}\lambda_{r}}\,\mid\ell\mid+c_{4}
\end{equation}
The arbitrary constant $c_{4}$ is null by imposing $\ell=0$ for
$R=R_{M}$. We thus obtain:

\begin{equation}
R\simeq R_{M}(1-\frac{f(f+2)}{2\lambda_{r}^{2}}\ell^{2})
\end{equation}

\begin{equation}
\frac{1}{R^{2}}\simeq\frac{1}{R_{M}^{2}}\frac{1}{1-\frac{f(f+2)}{\lambda_{r}^{2}}\ell^{2}}\simeq\frac{1}{R_{M}^{2}}(1+\frac{f(f+2)}{\lambda_{r}^{2}}\ell^{2})
\end{equation}
Therefore, from eq. (55), the quantum potential around the origin
can be approximated by the potential of a harmonic oscillator:

\begin{equation}
V_{Q}\simeq m\gamma_{o}c^{2}f\,(1+\frac{f(f+2)}{\lambda_{r}^{2}}\ell^{2})
\end{equation}

By replacing it into the equation of motion in $\tau$ (42) and imposing
$\ell(\tau=0)=0$ , we finally get the solutions referred to the motion
not directly observable:

\begin{equation}
\ell(0,0,\tau)\simeq-A\,sin(\frac{c}{\lambda_{r}}\,f\sqrt{2(f+2)})\tau)
\end{equation}

\begin{equation}
v_{i}(\tau)=-\frac{\partial\ell}{\partial\tau}\simeq A\,\frac{c}{\lambda_{r}}\,f\sqrt{2(f+2)})\,cos(\frac{c}{\lambda_{r}}\,f\sqrt{2(f+2)})\tau)
\end{equation}

The angular frequency obtained is more general than that one of the
classical \emph{Zitterbewegung}. This latter, $\frac{2m\gamma_{o}c{}^{2}}{\hbar}$,
we can get for $A=\frac{\hbar}{2m\gamma_{o}c}$ and $v_{i_{MAX}}=c$.
With these values, we also obtain: $\gamma_{o}\,f\sqrt{2(f+2)}=2$,
that, for $\gamma_{o}=1$, provides: $f\simeq0.839$.

However, based on eq. (55), dependence of \emph{f} on $\gamma_{o}$
appears forced in our model. By only imposing $v_{i_{MAX}}=c$, we
get $A=\frac{\lambda_{r}}{f\sqrt{2(f+2)}}$, where we will assume
that \emph{f }is of the order of unity. Recalling that these values
constitute the uncertainties on the corresponding observable quantities,
let us reconsider the uncertainty principle:

\begin{equation}
\frac{\lambda_{r}}{f\sqrt{2(f+2)}}\times m\gamma_{o}c=\frac{\hbar}{f\sqrt{2(f+2)}\gamma_{o}}\sim\frac{\hbar}{\gamma_{o}}
\end{equation}

\emph{As a novelty, there is the Lorentz factor in the denominator.
}Looking at eq. (44) this seems correct: there is a $\gamma_{o}$
which increases the relativistic mass and \emph{another} $\gamma_{o}$
which decreases the influence of $\hbar$ . At speeds close to \emph{c}
the quantum effects in the direction of motion must be negligible.

The maximum value of the quantum potential, $V_{QM}$, can be found
by applying energy conservation to the motion in $\tau$:

\begin{equation}
\frac{1}{2}m\gamma_{o}c^{2}+m\gamma_{o}c^{2}f=0+V_{QM}
\end{equation}
that provides: $V_{QM}=m\gamma_{o}c^{2}(f+\frac{1}{2})$; it coincides
with the \emph{purely quantum} energy $E_{Q}$.

$V_{Q}$, and so also \emph{H}, oscillates with an amplitude of $\frac{1}{2}m\gamma_{o}c^{2}$:
this value represents the quantum uncertainty in a measure of the
energy for a free particle with a minimum location given by $\frac{\lambda_{r}}{f\sqrt{2(f+2)}}$.
Let's find again the uncertainty principle by multiplying by the half-period
of the oscillation:

\begin{equation}
\frac{1}{2}m\gamma_{o}c^{2}\times\frac{\pi\lambda_{r}}{f\sqrt{2(f+2)\,}c}=\frac{\pi\hbar}{2\,f\sqrt{2(f+2)}\gamma_{o}}\sim\frac{\hbar}{\gamma_{o}}
\end{equation}
In correspondence of $V_{QM}$ we obtain the minimum value for \emph{R}:

\begin{equation}
R_{m}=\frac{R_{M}}{\sqrt{1+\frac{1}{2f}}}
\end{equation}

For a numerical estimate, we will now impose $A=\frac{\lambda_{r}}{2}$,
that is $f\sqrt{2(f+2)}=2$ $\Rightarrow f\simeq0.839$, obtaining:
$R_{m}\simeq0.79\,R_{M}$. Yet, using eq. (58) for an estimate of
the maximum value of $\ell$ we have:

\begin{equation}
\ell_{M}\simeq\lambda_{r}\,\sqrt{f\,(1-\frac{1}{\sqrt{1-\frac{1}{2f}}})}\simeq0.42\lambda_{r}
\end{equation}
close to $\frac{\lambda_{r}}{2}$, where the velocity is zero based
on the harmonic solution. 

Finally, a normalization of $R{}^{2}$ (numerical, given the non-simplification
by means of standard functions of (56)), can determine $R_{M}$.

\subsection{\emph{Zitterbewegung} in arbitrary direction}

We have got the intrinsic motion in the direction of motion $\hat{v}_{o}=\hat{x}$.
The generalization in an arbitrary direction $\hat{s}$, which forms
an angle \emph{$\vartheta$} with $\hat{v}_{o}$, is obtained by considering
the variable $\ell_{s}\equiv s-v_{s}\,t-\intop_{0}^{\tau}v_{is}(\zeta)\,d\zeta$,
where $v_{s}=v_{o}\,cos\vartheta$. Eqs. (44) and (45) are generalized
by:
\begin{equation}
H=\sqrt{m^{2}\gamma_{o}^{2}c^{4}-\hbar{}^{2}\frac{c^{2}}{\gamma_{s}^{2}}\frac{\ddot{R}}{R}}
\end{equation}

\begin{equation}
v_{s}\,\frac{\dot{R}}{R}\,(m\,\gamma_{o}\,c^{2}-H)+\frac{1}{2}\frac{\partial H}{\partial t}=0
\end{equation}
Where the dot denots derivative over $\ell_{s}$. Placing: $\lambda_{r}^{2}\equiv\frac{\hbar{}^{2}}{m^{2}c^{2}\gamma_{o}^{2}\gamma_{s}^{2}}$,
we get back the same equations we have considered. The Hamiltonian
and the quantum potential, given by (51) and (55), do not change,
not even the amplitude of their oscillations. By imposing the equation
of motion: 

\begin{equation}
m\gamma_{o}\frac{dv_{is}}{d\tau}=\dot{V}_{Q}
\end{equation}
we obtain an intrinsic harmonic motion described again by (61) and
(62), but with the new value of $\lambda_{r}$. The consequence on
the uncertainty principle - equations (63) and (65) - is the presence
of $\gamma_{s}$ instead of $\gamma_{o}$ in the denominator. Therefore,
in the particular case of \emph{$\hat{s}$ }perpendicular to\emph{
$\hat{v}_{o}$}, $\gamma_{s}$=1 and the standard uncertainty principle,
independent of speed, is re-established.

\subsection{Non-relativistic limit; singularity in $v_{o}=0$ }

For the non-relativistic ($v_{o}\ll c$) and $v_{o}\rightarrow0$
cases it is sufficient to approximate $\gamma_{o}$ in the previous
equations (respectively with $1+\frac{v^{2}}{2c^{2}}$ and 1), obtaining
still high frequency oscillatory motions in $\tau$. These results
cannot be obtained starting from the Schrödinger equation because
from it, for a $\psi$ described by eq. (1), the non-relativistic
approximation of eq. (69) cannot be deduced. 

In $v_{o}=0$ there is a singularity; in fact, from equation (45)
we have that\emph{ H}, and therefore $V_{Q}$, is constant. So, $v_{i}=const$,
contrary to the case $v_{o}\rightarrow0$. 

Finally, we see how it \emph{would} be possible to achieve the non-relativistic
potential of Bohm (6) from eq. (68), or:

\begin{equation}
H=m\gamma_{o}c^{2}\,\sqrt{1-\lambda_{r}^{2}\,\frac{\ddot{R}}{R}}
\end{equation}

with $\lambda_{r}^{2}\equiv\frac{\hbar{}^{2}}{m^{2}c^{2}\gamma_{o}^{2}\gamma_{s}^{2}}$
. The first step is to consider small the second addend in the square
root; but this approximation \emph{is not} non-relativistic: if true,
it holds - \emph{a fortiori} - also for relativistic speeds\footnote{Often it is said, incorrectly, that this approximation is obtained
for $c\rightarrow\infty.$ }. One would get:

\begin{equation}
H\simeq m\gamma_{o}c^{2}\,(1-\frac{\hbar{}^{2}}{2m^{2}c^{2}\gamma_{o}^{2}\gamma_{s}^{2}}\frac{\ddot{R}}{R})
\end{equation}

and finally, for $v_{o}\ll c$, and taking into account the eq. (8):

\begin{equation}
H=mc^{2}+\frac{1}{2}mv^{2}-\frac{\hbar{}^{2}}{2m}\,\frac{\nabla{}^{2}R}{R}
\end{equation}

in agreement with eq. (5).

However, \emph{this approximation is never legitimate for a free particle}.
In fact, in this case it has been found that the maximum value of
the quantum potential is \emph{always} at least of the order of $m\gamma_{o}c^{2}$:
it does not depend on $\hbar$! So the second addend in the square
root is \emph{never} small compared to the first one. What happens
as the rest mass increases is that the amplitude of the motion's oscillation
decreases, until it becomes negligible for sufficiently large masses.

\subsection{Standard \emph{Zitterbewegung}}

The standard \emph{Zitterbewegung} can be obtained interpretating
the oscillatory motion in $\tau$ as occurring in \emph{t}. First,
in the standard view \emph{R} is considered time independent, so the
second side of eq. (19) is simply $-\frac{H^{2}}{\hbar{}^{2}}$, with
$H=const$. In the new theory this means demanding $\frac{1}{R}\frac{\partial^{2}R}{\partial t^{2}}-2\,\frac{i}{\hbar}\,\frac{H}{R}\frac{\partial R}{\partial t}=0$,
that provides $R\propto\,e^{\frac{2i}{\hbar}H\,t}$ and thence $V_{Q}\propto e^{-\frac{4i}{\hbar}H\,t}$.
Therefore, referring eq. (42) at \emph{t} time and conveniently determining
the arbitrary constants, one gets $v_{i}\propto c\,e^{-\frac{2i}{\hbar}H\,t}$,
as found in the Heisenberg \emph{picture}.

\section{Final remarks}

The new Bohmian model is obtained by introducing a new independent
temporal dimension, $\tau$, whose relative movements are not directly
observable but constitute the uncertainties themselves of all the
observables of quantum mechanics. Unlike Bohm's original theory, the
quantum potential does not affect the observable motion in\emph{ t},
as occurs for a normal external potential, but it only determines
the intrinsic one in $\tau$. This peculiarity can be understood considering
the \textquotedbl{}retroactive\textquotedbl{} nature of the quantum
potential, which originates from the same wave function of the particle.

The intrinsic motion found for a free particle, so, is due to an interaction
of the particle with its own wave (without the need to invoke the
antiparticle) and although is more general than the \emph{Zitterbewegung}
discovered by Schrödinger, can be well approximated with it. This
movement is caused by the quantum potential and gives rise to the
uncertainty of a measure of the particle's position. The Hamiltonian
itself is not constant, but is a wave that follows the particle, function
of both temporal variables; the variability in $\tau$ causes an oscillation
which, like for any other observable, produces its indeterminacy.

The evanescence of the uncertainty principle in the direction of motion,
for velocities close to\emph{ c}, becomes clear just by observing
the general equation (68): in it, not only we have a Lorentz factor
that multiplies the rest mass, but also another one - equal to 1 only
in perpendicular direction to the direction of motion - which directly
reduces the quantum potential, which is the cause of motion in $\tau$
and hence of quantum uncertainties.

The consideration of a new independent time parameter is an unusual
and certainly not intuitive idea; but its mathematical simplicity
is disarming. Traveling in an independent time dimension, a particle
or photon can instantly (in relation to the usual time) interact with
more slits (as in a Davisson-Germer diffraction), it can run across
all the possible paths between two points, also exceeding the speed
of light (in relation to the usual time), it can admit a \emph{non-local}
relationship with another particle ... All things that \emph{it actually
does} \emph{based on experiments.}

In the studied model all these peculiarities are explained within
a deterministic picture, along with the prediction of a modification
of the uncertainty principle that could be tested in high-energy accelerators.

\end{document}